\documentclass[showpacs,preprintnumbers,amsmath,amssymb,superscriptaddress,11pt]{revtex4}

\usepackage{graphicx}
\usepackage{dcolumn}
\usepackage{bm}

\def\ra{\rangle}
\def\la{\langle}
\def\dag{^\dagger}

\begin{document}
\quad 
\vspace{-2cm}


\vspace*{0.5cm}

\title{Entanglement in an SU($n$) Valence-Bond-Solid State}

\author{Hosho Katsura}
\email{katsura@appi.t.u-tokyo.ac.jp}
\affiliation{Department of Applied Physics, University of Tokyo,
7-3-1, Hongo, Bunkyo-ku, Tokyo 113-8656, Japan}

\author{Takaaki Hirano}
\email{hirano@pothos.t.u-tokyo.ac.jp}
\affiliation{Department of Applied Physics, University of Tokyo,
7-3-1, Hongo, Bunkyo-ku, Tokyo 113-8656, Japan}

\author{Vladimir E. Korepin}
\email{korepin@insti.physics.sunysb.edu}
\affiliation{C.N. Yang Institute for Theoretical Physics, State
University of New York at Stony Brook, Stony Brook, NY 11794-3840,
USA}

\date{\today}

\begin{abstract}
We investigate entanglement properties in the ground state of the open/periodic SU($n$) generalized valence-bond-solid state consisting of representations of SU($n$). We obtain exact expression for the reduced density matrix of a block of contiguous spins and explicitly evaluate the von Neumann and the R\'enyi entropies. We discover that the R\'enyi entropy is independent of the parameter $\alpha$ in the limit of large block sizes and its value $2 \log n$ coincides with that of von Neumann entropy. We also find the direct relation between the reduced density matrix of the subsystem and edge states for the corresponding open boundary system. 
\end{abstract}

\maketitle
\section{Introduction}
There is  considerable current interest in quantifying entanglement
in various quantum many-body systems. Entanglement in spin chains, correlated electrons, interacting bosons
and other  models was studied in detail
\cite{OAFF,ON,VLRK,LRV,JK,K,ABV,VMC,PP,FS,V2,keat,xy,salerno, zanardi, zanardi3, 
honk1,honk2, kais2,Eisert}.
Entanglement is a resource for quantum computation, it shows how much 
correlation we can use to control quantum devices \cite{BD,L}. 
There are several different measures of entanglement. The most famous is the von Neumann (entanglement) entropy of a subsystem \cite{BD}. 
This measure has recently been used to detect quantum phase transitions and topological/quantum order \cite{Levin,Kitaev} in strongly correlated systems.
We can also use the R\'enyi entropy to quantify the entanglement. 
The Renyi entropy was first proposed in information theory \cite{Renyi}. 
The von Neumann entropy $S(\rho_A)$ and the Renyi entropy $S_{\alpha}(\rho_A)$ are defined as follows: 
\begin{eqnarray}
S(\rho_A)&=&-{\rm Tr}(\rho_A \log\rho_A)\qquad  
\label{edif}\\
S_{\alpha}(\rho_A)&=&\frac{1}{1-\alpha}\log {\rm Tr}(\rho_A^{\alpha}),
\qquad \alpha\neq 1 ~~\textrm{and}~~ \alpha>0.  \qquad  
\label{olds}
\end{eqnarray}
Here $\rho_A$ is the reduced density matrix of subsystem $A$ and the power $\alpha$ is an arbitrary parameter. 
The R\'enyi entropy characterizes the mixed state much better: if we know the R\'enyi entropy at any $\alpha$ we know all eigenvalues of the density matrix.
  
  Studying entanglement also helps to understand the physics of quantum spin systems  \cite{OAFF,ON}. The model introduced by  Affleck, Kennedy, Lieb, and Tasaki (AKLT model) \cite{AKLT,AKLT0} plays a very important role in condensed matter physics.
The exact ground state of this model is known as the Valence-Bond Solid (VBS) state. 
In 1983, Haldane \cite{H} conjectured that the antiferromagnetic Heisenberg Hamiltonian describing  half-odd-integer spins is gap-less, but for integer spins it has a gap. 
The AKLT model agrees with this conjecture and enables us to understand the ground-state properties of gapped spin chains in a unified fashion. 
The construction of the AKLT-type model is not restricted to one dimension. 
In fact, the AKLT model was formulated on an arbitrary graph and an integration over classical spins was used for the evaluation of correlation functions in the VBS ground state \cite{KK}.
The VBS state has attracted revived interest from the viewpoint of quantum information theory.  
An implementation of the AKLT model in optical lattices was proposed recently \cite{GMC}, and the use of the AKLT model for {\it universal quantum computation} was discussed in \cite{VC}. The VBS state is also closely related to the Laughlin wave function \cite{L0} and to the fractional quantum Hall effect \cite{AAH}. 
Entanglement in $S=1$ AKLT model was first considered in \cite{Fan}. Then the results were generalized to the arbitrary integer spin case in \cite{Katsura}. 
An interesting generalization of the AKLT model to the SU($n$) version was constructed in \cite{Greiter_su3,GreiterYoung}.

In this paper we study entanglement in SU($n$) version of the AKLT model.
We consider entanglement of a block of spins with the rest of the ground state.
We evaluate the von Neumann and the R\'enyi entropies of the block. 
We first confirm that the von Neumann entropy of a large block of spins reaches saturation. 
This is a partial proof of the conjecture proposed by Vidal {\it et al} \cite{VLRK}. 
In the SU(2) AKLT and the XY spin chains, this conjecture has already been  proved and the limiting entropy of the large block of spins was explicitly calculated \cite{Fan,Katsura,xy,xy2,xy3,xy4}.  
We find the essential simplification in the limit of large block sizes.
In this limit, we discovered that the R\'enyi entropy  is independent of $\alpha$, actually it coincides with the value of von Neumann entropy $2\ln n$. 
This means  that the density matrix of the block is proportional to identical matrix of dimension $n^2$. It is much different form XY spin chain, where the density matrix of the large  block is infinite dimensional and eigenvalues are different (see \cite{ikf}). 
We also explore the connection between the reduced density matrix of the subsystem and  degenerate ground states for the corresponding open boundary system.

The paper is organized as follows. In the next section, we will study entanglement in the ground state of the SU($n$) AKLT model with boundary spins. This section is the main part of this paper. 
The von Neumann entropy and the R\'enyi entropy will be evaluated for the block of neighboring spins. 
We will also discuss the direct relation between the reduced density matrix for the block and the degenerate ground states of the AKLT model with an open boundary condition. 
In the third section, we will investigate entanglement in the ground
state of the SU($n$) AKLT model with a periodic boundary condition. In
this section, we will obtain the explicit form of the reduced density
matrix. 
The finite size effect of 
the von Neumann and the R\'enyi entropies will be studied. 
The last section will be devoted to summary and discussions. 

\section{SU($n$) VBS state with boundary spins}
\subsection{Construction of SU($n$) VBS state}
\begin{figure}
\includegraphics[width=15cm,clip]{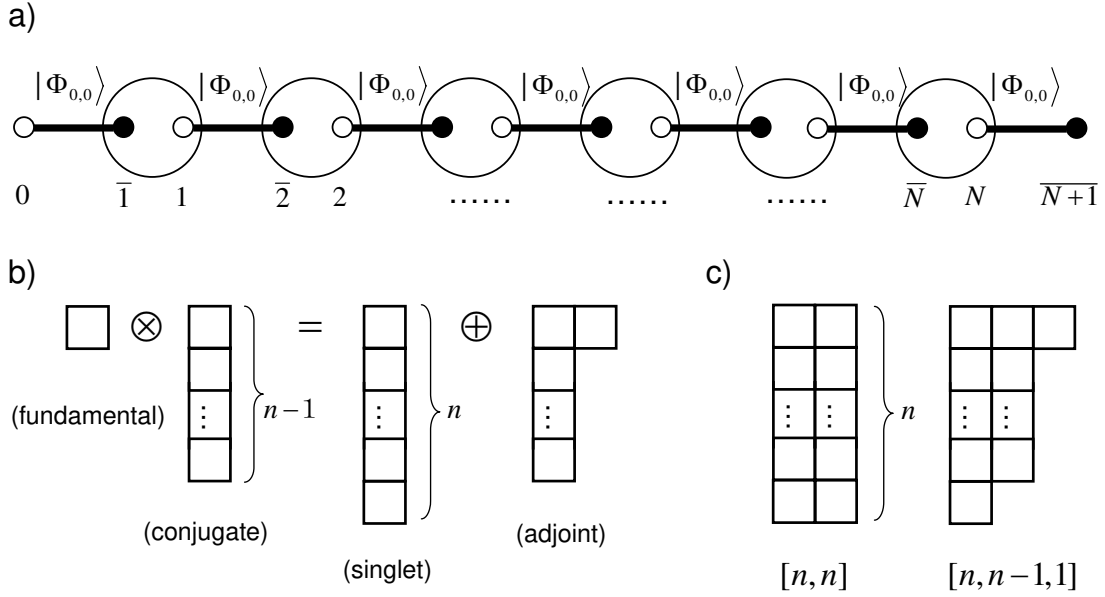}
\caption{a)Construction of the SU($n$) VBS solid state. White and black dots represent the SU($n$) fundamental and its conjugate representations, respectively. A dimer corresponds to the singlet state $|\Phi_{0,0}\ra$ and 
a circle denotes the projection onto the adjoint representation. 
b) The decomposition rule for the tensor product of $\square \otimes {\bar \square}$. c)Young tableaux corresponding to [$n,n$] and [$n,n-1,1$]}
\label{construction}
\end{figure}

In this section, we consider the SU($n$) VBS state with boundary spins and calculate the von Neumann entropy (entanglement entropy) and the R\'enyi entropy of a block of contiguous spins. What we mean by "spin" in our system is an adjoint representation of SU($n$). 
We shall first construct an SU($n$) VBS state which consists of $N$ adjoint representation of SU($n$) in the bulk and fundamental and conjugate representations of SU($n$) on the boundary. First, we prepare sites $k$ ($k=0, 1, ..., N$) and ${\bar k}$ ($k=1, 2, ..., N+1$) and 
arrange SU($n$) singlets consisting of a fundamental ($\square$) and its conjugate ($\bar {\square}$) representations as shown in Fig. (\ref{construction}.a) \cite{Zohar}. 
We assign $|j\ra$ $(j=0,1,2, ..., n-1)$ to the fundamental representation, while $|{\bar{j}}\ra$ $(j=0,1,2, ..., n-1)$ to the conjugate representation. $|{\bar j}\ra$ can be represented by the tensor product of $(n-1)$ $|j\ra$s as
\begin{equation}
|\bar{j}\ra \equiv \frac{1}{\sqrt{(n-1)!}}\sum_{\alpha_2, ..., \alpha_n} \epsilon^{j \alpha_2,...,\alpha_n}|\alpha_2,...,\alpha_n \ra
\label{conjugate}
\end{equation}
where $\epsilon^{j \alpha_2,...,\alpha_n}$ is a totally antisymmetric tensor of rank $n$. Using $|j\ra$ and $|{\bar j}\ra$, an SU($n$) singlet state $|\Phi_{0,0}\ra$ can be represented as a maximally entangled state:
\begin{equation}
|\Phi_{0,0}\ra = \frac{1}{\sqrt{N}}\sum_{j=0}^{N-1}|j\ra |\bar{j}\ra.
\end{equation}
The above relation can be easily confirmed by inserting the resolution of the identity $1=\sum_{j=0}^{n-1}|j\ra \la j|$ and substituting Eq.(\ref{conjugate}). Next, we prepare the adjoint representation of SU($n$) by projecting the tensor product $\square \otimes {\bar \square}$ onto an $(n^2-1)$-dimensional subspace. This procedure corresponds to circles in Fig. \ref{construction}.a). In Fig. \ref{construction}.b), we visualize the decomposition rule $\square \otimes \bar{\square}=({\rm singlet}) \oplus ({\rm adjoint})$. 
Then we obtain the SU($n$) adjoint representation at each composite site ($k,{\bar k}$). Henceforth we shall call this composite site $k$. 
Finally, we can represent the SU($n$) generalized VBS state as
\begin{equation}
|G\ra = (\otimes_{k=1}^N P_{k{\bar k}})|\Phi_{0,0}\ra_{0{\bar 1}} |\Phi_{0,0}\ra_{1{\bar 2}} \dots |\Phi_{0,0}\ra_{N\overline{N+1}},
\label{gs}
\end{equation}
where $P_{k{\bar k}}$ is a projection operator onto an adjoint representation of SU($n$). 

Let us now consider the Hamiltonian whose ground state is the above state we have constructed. The Hamiltonian can be constructed along the same line as the AKLT model:
\begin{equation}
H=\sum_{k=1}^{N-1} H_k + \pi_{0,1}+ \pi_{N,N+1}, \hspace{5mm}H_k=\sum_{Y} a_Y P^Y_{k,k+1},
\label{Ham}
\end{equation}
where $Y$ is a Young tableau which is neither [$n, n$] nor [$n$, $n-1$, 1]. 
Here we have assigned [$\kappa_1, ..., \kappa_{\lambda_1}$] to the Young tableau $Y$, where 
$\kappa_j$ is the number of boxes in the $j$-th column and $\lambda_1$ is the number of boxes in the first row.
$P^Y_{k,k+1}$ is a projection operator which projects (adjoint)$\otimes$(adjoint) onto a representation characterized by $Y$ and the coefficient $a_Y$ can be an arbitrary positive number. 
The reason why [$n, n$] and [$n$, $n-1$, 1] are excluded from the sum is following. 
Since $\square$ at site $k$ and $\bar \square$ at site $k+1$ have already formed a singlet in the ground state (\ref{gs}), the possible representations obtained from the decomposition of (adjoint)$\otimes$(adjoint) are restricted to [$n, n$] and [$n$, $n-1$, 1] (graphically shown in Fig.1.c)).  
$\pi_{0,1}$ and $\pi_{N,N+1}$ are boundary terms which assure the uniqueness of the ground state of this Hamiltonian. $\pi_{0,1}$ and $\pi_{N,N+1}$ can be written in terms of the projection operators acting on the tensor products (fundamental)$\otimes$(adjoint) and (conjugate)$\otimes$(adjoint), respectively. 
By construction, the SU($n$) VBS state (\ref{gs}) is a zero-energy ground state of this Hamiltonian. 
We should note here that another construction of Hamiltonian by Greiter and Rachel\cite{GreiterYoung} is similar but slightly different from ours. 

\subsection{Reduced density matrix and von Neumann and R\'enyi entropies}
Next, we consider the reduced density matrices of subsystems of the ground state $|G\ra$. To calculate the reduced density matrix, it is more convenient to recast the chain of singlets $|\Phi_{0,0}\ra_{0{\bar 1}} |\Phi_{0,0}\ra_{1{\bar 2}} \dots |\Phi_{0,0}\ra_{N\overline{N+1}}$ in Eq.(\ref{gs}) in a different form. Let us first consider a chain of two singlets $|\Phi_{0,0}\ra_{0{\bar 1}} |\Phi_{0,0}\ra_{1{\bar 2}}$. We can rewrite this product state as
\begin{equation}
|\Phi_{0,0}\ra_{0{\bar 1}} |\Phi_{0,0}\ra_{1{\bar 2}}=
\frac{1}{n}\sum_{l=0}^{n-1}\sum_{m=0}^{n-1}|\Phi_{l,m}\ra_{0{\bar 2}} 
|\Phi_{l,-m}\ra_{{\bar 1}1},
\label{important_relation}
\end{equation}
where $|\Phi_{m,n}\ra$ is a basis of the maximally entangled state defined by
\begin{equation}
|\Phi_{l,m}\ra = (U_{l,m}\otimes I)|\Phi_{0,0}\ra.
\end{equation}
Here $I$ is an $n$-dimensional identity matrix and $U_{l,m}=X^lZ^m$ ($m,n=0, 1, ..., n-1$) are generalized Pauli matrices, where the unitary operators $X$ and $Z$ act on $|j\ra$ as $X|j\ra=|j+1 ({\rm mod}.n)\ra$ and $Z|j\ra=\omega^j|j\ra$ with $\omega=e^{2\pi i/n}$, respectively. 
One can easily show the relation (\ref{important_relation}) by using the fact that $|\Phi_{0,0}\ra$ is invariant under the action of $(U_{l,m}\otimes U_{l,-m})$ \cite{Fan_Bell}. This procedure can be regarded as a multi-dimensional generalization of entanglement swapping. Repeatedly using the relation (\ref{important_relation}), we can generalize in a straightforward way to a chain of singlet states:
\begin{eqnarray}
|\Phi_{0,0}\ra_{0{\bar 1}} |\Phi_{0,0}\ra_{1{\bar 2}} \dots |\Phi_{0,0}\ra_{N\overline{N+1}}&=&\frac{1}{n^N}
\sum_{(l_1,m_1)}\cdots \sum_{(l_N,m_N)}|\Phi_{l_1,-m_1}\ra_{{\bar 1}1}\cdots |\Phi_{l_N,-m_N}\ra_{{\bar N}N} \nonumber \\
&&\times (U_{l_1,m_1}\cdots U_{l_N,m_N}\otimes I)|\Phi_{0,0}\ra_{0\overline{N+1}},
\label{singlet_product}
\end{eqnarray}
where $(m_k,n_k)$ ($k=1, 2, ..., N$) runs from (0,0) to $(n-1,n-1)$.
To obtain the ground state $|G\ra$ from (\ref{singlet_product}), we have to make a projection onto the subspace of adjoint representation at each site $k$. Since the decomposition rule $\square \otimes \bar{\square}=({\rm singlet}) \oplus ({\rm adjoint})$ and the fact that $|\Phi_{0,0}\ra$ is an SU($n$) singlet, the vector space of the adjoint representation is spanned by $|\Phi_{l,-m}\ra$ ($(l,m)\ne (0,0)$). Then the only thing to do is to omit the summation over $(l_k,m_k)=(0,0)$ in Eq.(\ref{singlet_product}). The SU($n$) generalized VBS state can be rewritten as:
\begin{equation}
|G\ra =\frac{1}{(n^2-1)^{N/2}}\sum_{(l_1,m_1)\atop \ne(0,0)}\cdots\sum_{(l_N,m_N)\atop \ne(0,0)}|\Phi_{l_1,-m_1}\ra_{{\bar 1}1}\cdots |\Phi_{l_N,-m_N}\ra_{{\bar N}N}
(U_{l_1,m_1}\cdots U_{l_N,m_N}\otimes I)|\Phi_{0,0}\ra_{0\overline{N+1}},
\end{equation}
where we have already normalized $|G\ra$ by the factor $1/(n^2-1)^{N/2}$. 

Let us now consider the reduced density matrix of a block of contiguous spins of length $L$. We first suppose that the block starts from site $k$ and stretches up to $k+L-1$, where $k\ge 1$ and $k+L-1 \le N$. The reduced density matrix is obtained by taking the trace over the sites $j=0,1, ..., k-1$ and $j=k+L, ..., N, \overline{N+1}$ as
\begin{eqnarray}
\rho_L &=&{\rm Tr}_{1,...,k-1,k+L,...,N,0,\overline{N+1}}|G\ra \la G|
\nonumber \\
&=&\frac{1}{(n^2-1)^N} \sum_{(l_1,m_1)\atop \ne(0,0)}\cdots\sum_{(l_{k-1},m_{k-1})\atop \ne(0,0)}\sum_{(l_{k+L},m_{k+L})\atop \ne(0,0)}\cdots\sum_{(l_N,m_N)\atop \ne(0,0)}
\sum_{(l_k,m_k)\atop \ne(0,0)}\sum_{(l_k',m_k')\atop \ne(0,0)}\cdots \sum_{(l_{L+k-1},m_{L+k-1})\atop \ne(0,0)}\sum_{(l_{L+k-1}',m_{L+k-1}')\atop \ne(0,0)}
\nonumber \\
&\times &|\Phi_{l_k,-m_k}\ra_{{\bar k}k}\la\Phi_{l_k',-m_k'}|\cdots|\Phi_{l_{L+k-1},-m_{L+k-1}}\ra_{\overline{L+k-1}L+k-1}\la \Phi_{l_{L+k-1}',-m_{L+k-1}'}| \nonumber \\
& \times &{\rm Tr}_{0,\overline{N+1}}(U_1 V U_2 \otimes I)|\Phi_{0,0}\ra_{0\overline{N+1}}\la \Phi_{0,0}|(U_1 V' U_2 \otimes I)\dag,
\label{RDM1}
\end{eqnarray}
where $U_1=U_{l_1,m_1}\cdots U_{l_{k-1},m_{k-1}}$, $U_2=U_{l_{L+k},m_{L+k}}\cdots U_{l_N,m_N}$, $V=U_{l_k,m_k}\cdots U_{l_{L+k-1},m_{L+k-1}}$ and $V'=U_{l_k',m_k'}\cdots U_{l_{L+k-1}',m_{L+k-1}'}$.
To rewrite Eq. (\ref{RDM1}), we use the following property of $|\Phi_{0,0}\ra$:
\begin{equation}
(S \otimes T)|\Phi_{0,0}\ra =(ST^t\otimes I)|\Phi_{0,0}\ra=(I\otimes TS^t)|\Phi_{0,0}\ra,
\end{equation}
where $S$ and $T$ are $n$-dimensional unitary operations acting on $|j\ra$ and $|{\bar j}\ra$, respectively, and the superscript $t$ denotes the transposition. Using this property and the cyclic property of the trace, we can simplify the last part of Eq. (\ref{RDM1}) as
\begin{eqnarray}
&&{\rm Tr}_{0,\overline{N+1}}(U_1 V U_2 \otimes I)|\Phi_{0,0}\ra_{0\overline{N+1}}\la \Phi_{0,0}|(U_1 V' U_2 \otimes I)\dag \nonumber \\
&=&{\rm Tr}_{0,\overline{N+1}}(U_1\otimes I)(V \otimes I)(U_2 \otimes I)|\Phi_{0,0}\ra_{0\overline{N+1}}\la \Phi_{0,0}|(U_2 \otimes I)\dag(V' \otimes I)\dag(U_1 \otimes I)\dag \nonumber \\
&=&{\rm Tr}_{0,\overline{N+1}}(V \otimes I)(I \otimes U_2^t)|\Phi_{0,0}\ra_{0\overline{N+1}}\la \Phi_{0,0}|(I \otimes U_2^t)\dag(V' \otimes I)\dag \nonumber \\
&=&{\rm Tr}_{0,\overline{N+1}}(I \otimes U_2^t)(V \otimes I)|\Phi_{0,0}\ra_{0\overline{N+1}}\la \Phi_{0,0}|(V' \otimes I)\dag(I \otimes U_2^t)\dag \nonumber \\
&=&{\rm Tr}_{0,\overline{N+1}}(V \otimes I)|\Phi_{0,0}\ra_{0\overline{N+1}}\la \Phi_{0,0}|(V' \otimes I)\dag.
\label{trace_relation}
\end{eqnarray}
Since (\ref{trace_relation}) does not depend on $(l_1,m_1), \cdots, (l_{k-1},m_{k-1})$ and $(l_{k+L},m_{k+L}), \cdots, (l_N,k_N)$, we can rewrite Eq. (\ref{RDM1}) as 
\begin{eqnarray}
\rho_L
&=&\frac{1}{(n^2-1)^L}
\sum_{(l_k,m_k)\atop \ne(0,0)}\sum_{(l_k',m_k')\atop \ne(0,0)}\cdots \sum_{(l_{L+k-1},m_{L+k-1})\atop \ne(0,0)}\sum_{(l_{L+k-1}',m_{L+k-1}')\atop \ne(0,0)}
|\Phi_{l_k,-m_k}\ra_{{\bar k}k}\la\Phi_{l_k',-m_k'}|\nonumber \\
&\cdots&|\Phi_{l_{L+k-1},-m_{L+k-1}}\ra_{\overline{L+k-1}L+k-1}\la \Phi_{l_{L+k-1}',-m_{L+k-1}'}| 
{\rm Tr}(V \otimes I)|\Phi_{0,0}\ra_{0\overline{N+1}}\la \Phi_{0,0}|(V' \otimes I)\dag.
\label{RDM2}
\end{eqnarray}
From the form of the reduced density matrix (\ref{RDM2}), we immediately notice that the reduced density matrix does not depend on both the starting site $k$ and the total length of the chain $N$. The same property for SU(2) $S=1$ VBS state has already been proved in Ref. \cite{Fan}. We can regard above result as an SU($n$) generalization of their result. 

Since the reduced density matrix is independent of both $k$ and $N$, we can set $N=L$ without loss of generality. We can further reduce the original  problem to that of the reduced density matrix of two end spins ($\square$ and $\bar \square$) using the following property of a bipartite pure state. Suppose that $|\Psi\ra_{AB}$ is a bipartite pure state of a total system $AB$. Then there exist orthonormal states $|\psi_j\ra_A$ for the subsystem $A$, and orthonormal states $|\phi_j\ra_B$ for $B$ such that 
\begin{equation}
|\Psi\ra_{AB}=\sum_j \sqrt{p_j}|\psi_j\ra_A |\phi_j\ra_B,
\label{Schmidt}
\end{equation}
where $p_j(>0)$ satisfy $\sum_j p_j=1$. This decomposition is called the Schmidt decomposition. The proof of the above theorem using the singular value decomposition can be found in Ref. \cite{Nielsen}. From Eq. (\ref{Schmidt}), one can immediately notice that 
the set of eigenvalues of $\rho_A={\rm Tr}_B |\Psi\ra_{AB}\la \Psi|$ coincides with that of $\rho_B={\rm Tr}_A |\Psi\ra_{AB}\la \Psi|$. 

Now we can reduce the eigenvalue-problem of $\rho_L$ to that of the reduced density matrix for end two spins $\rho_{\hat L}$. $\rho_{\hat L}$ has the following form:
\begin{equation}
\rho_{\hat L}=\frac{1}{(n^2-1)^L}\sum_{(l_1,m_1)\atop \ne(0,0)}\cdots \sum_{(l_L,m_L)\atop \ne(0,0)} (U \otimes I)|\Phi_{0,0}\ra_{0,\overline{L+1}}\la \Phi_{0,0}|(U \otimes I)\dag,
\label{edge_RDM}
\end{equation}
where $U=U_{l_1,m_1}\cdots U_{l_L,m_L}$. To evaluate the eigenvalues of $\rho_{\hat L}$, it is convenient to formulate the action of $(U_{l,m}\otimes I)$ as a transfer matrix. Let us first see the action of $(U_{l',m'}\otimes I)$ on a state $|\Phi_{l,m}\ra$:
\begin{equation}
(U_{l',m'}\otimes I)|\Phi_{l,m}\ra
=(X^{l'}Z^{m'}X^{l}Z^{m}\otimes I)|\Phi_{0,0}\ra
=\omega^{m'l}|\Phi_{l+l',m+m'}\ra,
\end{equation}
where we have used the relation $ZX=\omega XZ$.
Using the above relation, we can prove that 
\begin{equation}
(U_{l',m'}\otimes I)|\Phi_{l,m}\ra\la\Phi_{l,m}|(U_{l',m'}\otimes I)\dag =|\Phi_{l+l',m+m'} \ra\la \Phi_{l+l',m+m'}|.
\end{equation}
Next, we assign the vector 
$(0, ..., 0,1((l,m)$-th entry), 0, ..., 0)$^t$ to the state $|\Phi_{l,m}\ra \la\Phi_{l,m}|$. This one to one correspondence plays an essential role in our analysis. 
From this bijection, the operation $ \sum_{(l',m')\ne(0,0)}(U_{l',m'}\otimes I)|\Phi_{l,m}\ra\la\Phi_{l,m}|(U_{l',m'}\otimes I)\dag$ can be written in terms of $(n^2 \times n^2)$-dimensional matrix as
\begin{equation}
T \equiv
\begin{array}{c}
\longleftarrow \hspace{2mm} n^2 \hspace{2mm} \longrightarrow \\
\left(\begin{array}{ccccc}
0 & 1 & 1 & \cdots & 1 \\
1 & 0 & 1 & \cdots & 1 \\
1 & 1 & 0 & \cdots & 1 \\
\vdots & \vdots & \vdots & \ddots & \vdots \\
1 & 1 & 1 & \cdots & 0 \\
\end{array}\right).
\end{array}
\end{equation}
This transfer matrix can be diagonalized by the following unitary matrix:
\begin{equation}
U_c=\frac{1}{n}\left(
\begin{array}{ccccc}
1 & 1 & 1 & \cdots & 1 \\
1 & \zeta & \zeta^2 & \cdots   & \zeta^{n^2-1} \\
1 & \zeta ^2 & \zeta^4 & \cdots & \zeta^{2(n^2-1)} \\
\vdots & \vdots & \vdots & \ddots & \vdots \\
1 & \zeta^{n^2-1} & \zeta^{2(n^2-1)} & \cdots & \zeta^{(n^2-1)^2}\\
\end{array}\right),
\end{equation}
where $\zeta={\rm exp}(2\pi i/n^2)$. Then we can obtain the explicit form of the reduced density matrix $\rho_{\hat L}$ as
\begin{eqnarray}
\rho_{\hat L}&=&\frac{1}{(n^2-1)^L}T^L (1, 0, ..., 0)^t 
=\frac{1}{(n^2-1)^L} U_c [{\rm diag}(n^2-1,-1, ..., -1)]^L U_c\dag (1, 0, ..., 0)^t \nonumber \\
&=& \frac{1}{n^2}(1+(n^2-1)p_n(L))|\Phi_{0,0}\ra_{0\overline{N+1}}\la\Phi_{0,0}| + \frac{1}{n^2}\sum_{(l,m)\ne(0,0)}(1-p_n(L))|\Phi_{l,m}\ra_{0\overline{N+1}}\la\Phi_{l,m}|,
\label{RDM3}
\end{eqnarray}
where we have used the relation, $1+\zeta^k+\zeta^{2k}+\cdots +\zeta^{(n^2-1)k}=0$ ($1\le k \le n^2-1$) and $p_n(L)=(\frac{-1}{n^2-1})^L$.
Substituting $n=2$ into Eq. (\ref{RDM3}), one can reproduce the result of the SU(2) S=1 VBS state obtained in Ref. \cite{Fan}. 

Let us now start the evaluation of the von Neumann and the R\'enyi entropies of a block of $L$ contiguous spins. 
First, we shall examine the von Neumann entropy of the block. 
From the Schmidt decomposition and the definition of the von Neumann entropy $S(\rho_L)=S(\rho_{\hat L})=-{\rm Tr}_{1,2,...,L}(\rho_{\hat L}\log \rho_{\hat L})$, we obtain
\begin{equation}
S(\rho_L)=2 {\rm log} n -\frac{1+(n^2-1)p_n(L)}{n^2}{\rm log} (1+(n^2-1)p_n(L))
-(n^2-1)\frac{1-p_n(L)}{n^2}{\rm log}(1-p_n(L))
\end{equation}
with $p_n(L)=(\frac{-1}{n^2-1})^L$.
Similarly to the SU(2) integer-$S$ VBS states \cite{Fan, Katsura} and the XY spin chains in the gapped regime \cite{xy,xy2,xy3,xy4}, 
$S(\rho_L)$ is bounded by $2 \log n$ in the limit of large block sizes ($L \to \infty$) and approaches to this value exponentially fast in $L$. This is a partial proof of the conjecture proposed by Vidal {\it et al.} \cite{VLRK}, that the von Neumann entropy of a large block of spins in gapped spin chains  shows saturation. 
Next we shall examine the R\'enyi entropy of our system. From the definition of the R\'enyi entropy $S_{\alpha}(\rho_L)=\frac{1}{1-\alpha}\log {\rm Tr}(\rho_L^{\alpha})$ ($\alpha \ne 1$,and $\alpha >0$),  
\begin{equation}
S_{\alpha}(\rho_L)=\frac{1}{1-\alpha}\log(\lambda_{0,0}(L)^\alpha + (n^2-1)\lambda_{l,m \ne 0,0}(L)^\alpha),
\end{equation}
where 
\begin{equation}
\lambda_{l,m}(L) =\left\{
\begin{array}{cc}
\frac{1}{n^2}(1+(n^2-1)p_n(L)), & (l,m)=(0,0)\\
\frac{1}{n^2}(1-p_n(L)) & (l,m)\ne(0,0).
\end{array}\right.
\label{lambda}
\end{equation}
Now we consider the limit of large block sizes, {\it i.e.}, $L \to \infty$. In this case, $\lambda_{l,m}$ become degenerate and great simplification occurs:
\begin{eqnarray}
S_{\alpha}(\rho_\alpha)&=&\frac{1}{1-\alpha}\log \Big(\big(\frac{1}{n^2}\big)^\alpha+(n^2-1)\big(\frac{1}{n^2}\big)^\alpha \Big) \nonumber \\
&=& \frac{1}{1-\alpha}\log(n^2)^{1-\alpha} \nonumber \\
&=& 2 \log n.
\end{eqnarray}
Now we notice that the R\'enyi entropy is independent of $\alpha$ and furthermore coincides with the von Neumann entropy. 
This means that the reduced density matrix of a large block is proportional to
$n^2$-dimensional  identity matrix. 
In other words, a sufficiently large block of neighbouring spins in our SU($n$) VBS ground state is maximally entangled with the rest of the chain. Finally, we consider the analytic
property of the R\'enyi entropy on the complex-$\alpha$ plane. 
In the large block limit, $S_\alpha(\rho_{\infty})$ is independent of $\alpha$ and hence $S_\alpha$ is completely analytic on $\alpha$-plane. 
On the other hand, if we consider the finite-size block, $S_\alpha(\rho_L)$ has branch cuts starting from $\alpha_{\rm bc}$. 
The branch points $\alpha_{\rm bc}$ are determined from the condition$\lambda_{0,0}(L)^\alpha+(n^2-1)\lambda_{l,m\ne0,0}(L)^\alpha=0$. 
The explicit value of $\alpha_{\rm bc}$ is given by 
\begin{equation}
\alpha_{\rm bc}=\frac{\pm(2m+1)\pi i+\log(n^2-1)}{\log(\lambda_{0,0}(L)/\lambda_{l,m \ne 0,0}(L))}
\label{branch}
\end{equation}
with $m \in {\bf Z}$. Since all the eigenvalues $\lambda_{l,m}$ become degenerate in the limit of large block sizes, the denominator of Eq.(\ref{branch}) becomes zero and hence $\alpha_{\rm bc}$ converge to the point at infinity. 
This is the reason why the R\'enyi entropy in this limit is completely analytic on the complex $\alpha$-plane. 
We also find the novel even-odd alternation of the real part of $\alpha_{\rm bc}$, {\it i.e.}, ${\rm Re}[\alpha_{\rm bc}]>0$ for even $L$, while ${\rm Re}[\alpha_{\rm bc}]<0$ for odd $L$. 

\subsection{Reduced density matrix as a projector onto the subspace of edge states}
As we have seen in the previous subsection, the von Neumann and R\'enyi entropies are obtained from the reduced density matrix for end two spins ($\rho_{\hat L}$). The obtained results indicate that the block of bulk spins is maximally entangled with the rest in the limit of large block sizes. 
The number of degrees of freedom in the subsystem can be counted from the von Neumann entropy as $n^2$. This number coincides with the number of edge states which are degenerate ground states of the AKLT model with an open boundary condition. In the case of SU(2) AKLT model, the close relation between the von Neumann entropy and the number of edge states has been extensively discussed \cite{Katsura, Hirano}. 
In this subsection, we elucidate the direct relation between the reduced density matrix (\ref{RDM2}) and the edge states. 
First, we consider the open boundary SU($n$) AKLT model with $L$ sites. 
The Hamiltonian is given by 
$H_{\rm open}=\sum_{k=1}^{L-1} H_k$.
\begin{figure}
\includegraphics[width=11cm,clip]{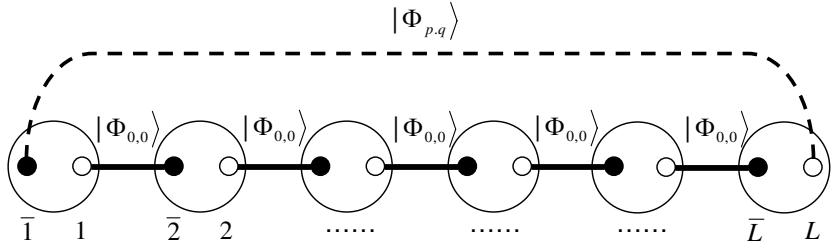}
\caption{Construction of the edge state $|p,q\ra$ in an open boundary spin
chain.
A white (black) dot represents the SU($n$) fundamental (conjugate)
representation. A circle denotes the projection onto the adjoint
representation. The dotted line corresponds to the state $|\Phi_{p,q}\ra$. }
\label{construction_open}
\end{figure}
The only difference from Eq.(\ref{Ham}) is that there are no boundary terms. 
The basis of the edge states is constructed as:
\begin{eqnarray}
|p,q\ra &\equiv& C_{p,q}\sum_{(l_1,m_1)\atop \ne(0,0)}\cdots \sum_{(l_L,m_L) \atop \ne(0,0)} |\Phi_{l_1,-m_1}\ra_{\bar 11}\cdots|\Phi_{l_{L-1},-m_{L-1}}\ra_{\overline{L-1}L-1} \nonumber \\
&\times& P_{L \bar L}((U_{p,q}U_{l_1,m_1}...U_{l_{L-1},m_{L-1}} \otimes I)|\Phi_{0,0}\ra_{L \bar L}),
\label{Edgestate}
\end{eqnarray}
where $C_{p,q}$ is a normalization factor. Any linear combination of (\ref{Edgestate}) is apparently the ground state of $H_{\rm open}$. 
The graphical representation of the construction of this state is shown in Fig. (\ref{construction_open}). The following orthogonality relation of edge state holds:
$ \la p,q| r,s \ra =C_{p,q}^2 (n^2-1)^{L}\delta_{p,r}\delta_{q,s}\lambda_{-p,-q}(L)$ 
(the subscripts $-p$ and $-q$ are modulo $n$).
This can be shown as follows:
\begin{eqnarray}
&&\la p,q| r,s \ra 
= C_{p,q} C_{r,s} \sum_{(l_1,m_1)\atop \ne(0,0)}\cdots\sum_{(l_{L-1},m_{L-1}) \atop \ne(0,0)} \nonumber \\
&\times&{}_{L \bar L}\la \Phi_{0,0}| (U_{p,q}U_{l_1,m_1}...U_{l_{L-1},m_{L-1}})\dag P_{L\bar L}(U_{r,s}U_{l_1,m_1}...U_{l_{L-1},m_{L-1}} \otimes I)|\Phi_{0,0}\ra_{L \bar L} 
\nonumber \\
&=& C_{p,q} C_{r,s}\sum_{(l,m)\atop \ne(0,0)} \sum_{(l_1,m_1)\atop \ne(0,0)} \cdots \sum_{(l_{L-1},m_{L-1}) \atop \ne (0,0)} {}_{L \bar L}\la \Phi_{0,0}| (U_{l_1,m_1}...U_{l_{L-1},m_{L-1}}\otimes I)\dag (U_{p,q}\otimes I)\dag |\Phi_{l,m}\ra_{L\bar L}
\nonumber \\
&\times & {}_{L \bar L}\la \Phi_{l,m}|
(U_{r,s}\otimes I)(U_{l_1,m_1}...U_{l_{L-1},m_{L-1}} \otimes I)|\Phi_{0,0}\ra_{L \bar L} 
\nonumber \\
&=& C_{p,q}C_{r,s}(n^2-1)^{L-1}\sum_{(l',m')} \lambda_{l',m'}(L-1) {}_{L \bar L}\la\Phi_{p+l'.q+m'}|(1-|\Phi_{0,0}\ra_{L \bar L}\la \Phi_{0,0}|)|\Phi_{r+l',s+m'}\ra_{L \bar L}
\nonumber \\
&=& C_{p,q}^2 (n^2-1)^{L}\delta_{p,r}\delta_{q,s}\lambda_{-p,-q}(L).
\end{eqnarray}
Here we have recalled Eq. (\ref{edge_RDM}) and have used the relation
\begin{eqnarray}
&& \sum_{(l_1,m_1)\atop \ne(0,0)} \cdots \sum_{(l_{L-1},m_{L-1}) \atop \ne (0,0)}(U_{l_1,m_1}...U_{l_{L-1},m_{L-1}} \otimes I)|\Phi_{0,0}\ra_{L \bar L}\la \Phi_{0,0}|  (U_{l_1,m_1}...U_{l_{L-1},m_{L-1}}\otimes I)\dagger\nonumber \\
&=&(n^2-1)^{L-1}\rho_{\widehat{L-1}}=(n^2-1)^{L-1}\sum_{(l,m)}\lambda_{l,m}(L-1)|\Phi_{l,m}\ra_{L\bar L}\la\Phi_{l,m}|.
\end{eqnarray}
The explicit form of the normalization factors $C_{p,q}$ are given by
$C_{p,q}= 1/\sqrt{(n^2-1)^{L}\lambda_{-p,-q}(L)}$.
Next, we try to write $\rho_L$ in terms of the basis of edge states. 
By the original definition, 
\begin{eqnarray} 
\rho_L &=&{\rm Tr}_{0,\overline{L+1}}|G\ra \la G| \nonumber \\
&=& \frac{1}{(n^2-1)^L}\sum_{(p,q)}\sum_{(l_1,m_1)\atop \ne(0,0)}\sum_{(l'_1,m'_1)\atop \ne(0,0)}\cdots \sum_{(l_{L-1},m_{L-1})\atop \ne (0,0)}\sum_{(l'_{L-1},m'_{L-1})\atop \ne (0,0)}
\nonumber \\
&\times& |\Phi_{l_1,-m_1}\ra \la \Phi_{l'_1,-m'_1}|
\cdots |\Phi_{l_{L-1},-m_{L-1}}\ra \la \Phi_{l'_{L-1},-m'_{L-1}}|
\nonumber \\
&\times& P_{L\bar L}(U_{p,q}U_{l_1,m_1}...U_{l_{L-1},m_{L-1}} \otimes I)|\Phi_{0,0}\ra_{L \bar L} \la \Phi_{0,0}|(U_{p,q}U_{l'_1,m'_1}...U_{l'_{L-1},m'_{L-1}} \otimes I)\dag P_{L \bar L}.
\end{eqnarray}
Then comparing with Eq. (\ref{Edgestate}), we obtain 
\begin{equation}
\rho_L=\sum_{(p,q)} \lambda_{-p,-q}(L) |p,q\ra \la p,q|,
\end{equation}
where $\lambda_{-p,-q}(=\lambda_{n-p,n-q})$  was defined in Eq. (\ref{lambda}). 
Therefore we can conclude that the reduced density matrix of a block of contiguous spins in the ground state is completely characterized by the ground states of corresponding open spin chain. 
In the limit of large block sizes, {\it i.e}., $L \to \infty$, $\rho_{L}$ can be written as
\begin{equation}
\rho_{L} = \frac{1}{n^2}\sum_{(p,q)}|p,q\ra \la p,q|.
\end{equation}
In this limit, the limiting density matrix can be regarded as a projector which projects on a subspace spanned by a degenerate ground states for open boundary AKLT model.  

\section{Periodic SU($n$) VBS state}
In this section, we shall focus on the von Neumann and the
R\'enyi entropies of the periodic SU($n$) VBS state. The periodic VBS state with length $N$ can be constructed by acting the projection operator on the edges of the VBS state with boundary spins.
We use the following representation for the ground state.
\begin{eqnarray}
|G_p\ra
&=&\frac{1}{\cal{N}}\sum_{(l_1,m_1)\atop \ne(0,0)}\cdots\sum_{(l_{L-1},m_{L-1})\atop \ne(0,0)}\sum_{(l_{L+1},m_{L+1})\atop \ne(0,0)}\cdots\sum_{(l_N,m_N)\atop\ne(0,0)}
\nonumber\\&\times&
|\Phi_{l_1,-m_1}\ra_{{\bar 1}1}\cdots
|\Phi_{l_{L-1},-m_{L-1}}\ra_{\overline{L-1}{L-1}}
|\Phi_{l_{L+1},-m_{L+1}}\ra_{\overline{L+1}L+1}\cdots
|\Phi_{l_{N},-m_{N}}\ra_{\overline{N}{N}}
\nonumber\\&\times&
P_{L\overline{L}}(U_{l_{L+1},m_{L+1}}\cdots U_{l_{N},m_N}U_{l_1,m_1}\cdots
U_{l_{L-1},m_{L-1}}\otimes I)|\Phi_{0,0}\ra_{L\overline{L}},
\end{eqnarray}
where ${\cal N}^2=(n^2-1)^N (1+(n^2-1)p_n(N))/n^2$ is a normalization factor. In the above state, the $L$-th site consists of the left ($\square$) and right ($\bar \square$) "spins" in the original open SU($n$) VBS state.  
The reduced density matrix of a block of contiguous spins with length
$L$ is given by
\begin{eqnarray}
\rho_{N,L}&=&{\rm Tr}_{L+1\cdots N} 
			   |G_p\rangle\langle G_p|
\nonumber\\
&=&\frac{1}{{\cal N}^2} 
\sum_{(l_1,m_1)\atop \ne(0,0)}\sum_{(l_1',m_1')\atop \ne(0,0)}\cdots \sum_{(l_{L-1},m_{L-1})\atop \ne(0,0)}\sum_{(l_{L-1}',m_{L-1}')\atop \ne(0,0)} 
\sum_{(l_{L+1},m_{L+1})\atop \ne(0,0)}\cdots \sum_{(l_{N},m_{N})\atop \ne(0,0)}\nonumber \\
&\times& |\Phi_{l_1,-m_1}\ra_{{\bar 1}1}\la\Phi_{l_1',-m_1'}|\cdots|\Phi_{l_{L-1},-m_{L-1}}\ra_{\overline{L-1}L-1}\la \Phi_{l_{L-1}',-m_{L-1}'}| \nonumber \\
&\times&P_{L,\overline{L}}(V_1U_3 \otimes I)|\Phi_{0,0}\ra_{L\overline{L}}\la \Phi_{0,0}|(V'_1U_3 \otimes I)\dag P_{L\overline{L}},
\label{rho_N_L}
\end{eqnarray}
where $V_1=U_{l_1,m_1}\cdots U_{l_{L-1},m_{L-1}}$,
$V_1'=U_{l_1',m_1'}\cdots U_{l_{L-1}',m_{L-1}'}$ and
$U_3=U_{l_{L+1},m_{L+1}}\cdots U_{l_{N},m_{N}}$.
Here it is convenient to introduce the following density matrix:
\begin{eqnarray}
\rho_{\widehat{
 N-L}}&=&\frac{1}{(n^2-1)^{N-L}}\sum_{(l_{L+1},m_{L+1})\atop \ne(0,0)}\cdots
 \sum_{(l_N,m_N)\atop \ne(0,0)} (U_3 \otimes
 I)|\Phi_{0,0}\ra_{L\overline{L}}\la \Phi_{0,0}|(U_3 \otimes I)\dag \nonumber \\
&=&\sum_{(l,m)}\lambda_{l,m}(N-L)|\Phi_{l,m}\ra_{L\overline{L}}\la\Phi_{l,m}|
\label{edge_RDM2}.
\end{eqnarray}
The above expression is obtained from Eq. (\ref{edge_RDM}) by replacing $L$ with $N-L$. 
Eq.(\ref{rho_N_L}) can be written in terms of $\rho_{\widehat{N-L}}$ as:
\begin{eqnarray}
\rho_{N,L}&=&\frac{1}{{\cal N}^2} (n^2-1)^{N-L}
\sum_{(l_1,m_1)\atop \ne(0,0)}\sum_{(l_1',m_1')\atop \ne(0,0)}\cdots \sum_{(l_{L-1},m_{L-1})\atop \ne(0,0)}\sum_{(l_{L-1}',m_{L-1}')\atop \ne(0,0)} 
|\Phi_{l_1,-m_1}\ra_{{\bar 1}1}\la\Phi_{l_1',-m_1'}|
\nonumber \\
\times&\cdots&|\Phi_{l_{L-1},-m_{L-1}}\ra_{\overline{L-1}L-1}\la \Phi_{l_{L-1}',-m_{L-1}'}| 
P_{L\overline{L}}(V_1 \otimes I) \rho_{\widehat{N-L}} (V'_1 \otimes I)^{\dag}P_{L\overline{L}},
\nonumber\\&=&
\frac{1}{{\cal N}^2} (n^2-1)^{N-L}
\sum_{(l,m)}\lambda_{l,m}(N-L)
\sum_{(l_1,m_1)\atop \ne(0,0)}\sum_{(l_1',m_1')\atop \ne(0,0)}\cdots \sum_{(l_{L-1},m_{L-1})\atop \ne(0,0)}\sum_{(l_{L-1}',m_{L-1}')\atop \ne(0,0)}
|\Phi_{l_1,-m_1}\ra_{{\bar 1}1}\la\Phi_{l_1',-m_1'}|
\nonumber \\
\times &\cdots&|\Phi_{l_{L-1},-m_{L-1}}\ra_{\overline{L-1}L-1}\la \Phi_{l_{L-1}',-m_{L-1}'}| 
P_{L\overline{L}}(V_1 \otimes I)
|\Phi_{l,m}\ra_{L\overline{L}}\la\Phi_{l,m}|
 (V'_1 \otimes I)^{\dag}P_{L\overline{L}}.
\end{eqnarray}
Here we recall the basis of edge states (\ref{Edgestate}) and obtain 
\begin{eqnarray}
\rho_{N,L}&=&\frac{1}{{\cal N}^2} (n^2-1)^{N-L}
\sum_{(l,m)} \frac{\lambda_{l,m}(N-L)}{C_{l,m}^2}|l,m\ra\la l,m|,
\nonumber\\&=&
\sum_{(l,m)} \lambda_{l,m}(N,L)|l,m\ra\la l,m|,
\end{eqnarray}
where $\lambda_{l,m}(N,L)=\frac{\lambda_{l,m}(N-L) }{{\cal
N}^2C_{l,m}^2}(n^2-1)^{N-L}$ are eigenvalues of the reduced density matrix.  
They are explicitly given by 
\begin{eqnarray}
\lambda_{l,m}(N,L)=
\left\{
\begin{array}{cc}
\frac{1}{n^2}\frac{(1+(n^2-1)p_n(N-L))(1+(n^2-1)p_n(L))}{(1+(n^2-1)p_n(N))}&(l,m)=(0,0)\\
\frac{1}{n^2}\frac{(1-p_n(N-L))(1-p_n(L))}{(1+(n^2-1)p_n(N))}
&(l,m)\ne(0,0)
\end{array}
\right.
\label{ev_open}
\end{eqnarray}
Putting $n=2$ into Eq. (\ref{ev_open}) reproduces the results of $S=1$ case obtained in Ref.[\cite{Hirano}]. 

Now we know the reduced density matrix of a block with arbitrary system size $N$, subsystem size $L$ and internal degrees of freedom $n$, we can study the finite size effect on entanglement properties.
In the limit of $N \to \infty$, the set of eigenvalues of the density matrix $\rho_{N,L}$ becomes equivalent to that of the VBS state with boundary spins, and hence the entanglement properties are identical to each other. 
The explicit form of the entanglement is similar to the one in previous
section. The von Neumann and the R\'enyi entropies for periodic case are explicitly written as 
\begin{eqnarray}
S(\rho_{N,L})&=&-\lambda_{0,0}(N,L)\log\lambda_{0,0}(N,L)-(n^2-1)\lambda_{l,m\neq 0,0}(N,L)\log\lambda_{l,m\neq 0,0}(N,L), \nonumber \\
S_\alpha(\rho_{N,L})&=&\frac{1}{1-\alpha}\log(\lambda_{0,0}(N,L)+(n^2-1)\lambda_{l,m\neq 0,0}(N,L)).
\end{eqnarray}
Similarly to the previous section, both $S(\rho_{N,L})$ and $S_\alpha(\rho_{N,L})$ approach $2\log{n}$ in the thermodynamic limit. 
Here what we mean by {\it thermodynamic limit} is the limit in which $N\rightarrow\infty,L\rightarrow\infty$ with parameter $L/N$ be fixed.

As we saw in the calculation of the reduced density matrix, 
we found non-trivial consequence that the reduced density matrix is completely
written in terms of the edge states of the open boundary AKLT model which
corresponds to the Hamiltonian describing the subsystem ripped off by tracing out. 
It clarifies the edge state interpretation of the entanglement entropy,
which is also discussed in several papers\cite{Ryu,Katsura,Hirano}, in more
detail.
We can see the separation of the eigenvalues of the reduced density
matrix, {\it i.e.}, $\lambda_{0,0}(N,L)\neq\lambda_{l,m\ne 0,0}(N,L)$ in finite
size systems. 
Here we shall explain the physical meaning of this separation. 
This is due to qualitative difference between the singlet state $|0,0\ra$ and the adjoint states
$|l,m\ra$, $(l,m)\neq(0,0)$ induced by the effective coupling of residual edge ``spins'' on the boundaries of the subsystem. 
In the thermodynamic limit, the reduced density matrix is proportional
to the identity matrix, which indicates the absence of the effective
coupling in this limit and hence the edge spins behave freely without interaction.
Therefore, the von Neumann entropy is proportional to the logarithm of the number of degerees of freedom due to the residual edge spins.

\section{Summary and discussions} \label{sec:summary}
We analyzed entanglement in the ground state of the  SU($n$) version of AKLT model.
We consider a block of spins in  the ground state, it is in the mixed state. 
We evaluated the von Neumann entropy and the R\'enyi entropy of the block. 
We first examined the VBS ground state with boundary spins. 
We found that the great simplification occurs in the limit of large block sizes.
In this case the R\'enyi entropy is independent of the parameter $\alpha$ and furthermore it coincides with the value of von Neumann entropy $2\ln n$. This means  that the density matrix of the block is proportional to identical matrix of dimension $n^2$. We clarified that subspace of eigenvectors of the density matrix with non-zero eigenvalues describes the degenerate ground state of the block, {\it i.e.}, edge states. 
We studied the finite size effect on the R\'enyi entropy in terms of analyticity on the complex $\alpha$-plane. 
Then we studied the periodic VBS state. In this case, we also obtained the exact expression for the reduced density matrix and found the essential simplification in the thermodynamic limit.

\section{Acknowledgements}
The authors are grateful to
S. Murakami, and Y. Hatsugai for fruitful discussions.
This work was supported in part by NSF grant DMS-0503712 (V.E.K.), Grant-in-Aids (Grant No. 15104006, No. 16076205, and No. 17105002) and NAREGI Nanoscience Project from the Ministry of Education, Culture, Sports, Science, and Technology. 
H.K. was supported by the Japan Society for the Promotion of Science.

\end{document}